\documentstyle[preprint,aps]{revtex}

\input psfig

\newcommand{\Et}{\mbox{$E_{\rm T}$}}
\newcommand{\Etmax}{\mbox{$E_{\rm T}^{\rm max}$}}
\newcommand{\Etjet}{\mbox{$E_{\rm T}^{\rm jet}$}}
\newcommand{\modeta}{\mbox{$\mid \! \eta  \! \mid$}}
\newcommand{\als}{\mbox{${\alpha_{{\scriptscriptstyle S}}}$}}
\newcommand{\pbarp}{\mbox{$\bar{p}p$}}

\begin{document}

\preprint{D\O\ Note 3285, \vspace{5mm} FERMILAB-Pub-98/017}

\title{An Investigation of Uncertainties in the QCD NLO Predictions of 
the Inclusive Jet Cross Section in \pbarp\ Collisions at $\sqrt{s}$ =
1.8 TeV and 630 GeV}

\author{
B.~Abbott,$^{5}$ 
I.A.~Bertram,$^{6}$ 
M.~Bhattacharjee,$^{7}$ 
G.~Di~Loreto,$^{3}$
V.D.~Elvira,$^{7}$ 
R.~Hirosky,$^{2}$
T.~Joffe-Minor,$^{6,\ast}$
J.~Krane,$^{4}$ 
and F.~Nang,$^{1}$
}
\address{
\centerline{$^{1}$University of Arizona, Tucson, Arizona 85721}
\centerline{$^{2}$Florida State University, Tallahassee, Florida 32306}
\centerline{$^{3}$Michigan State University, East Lansing, Michigan 48824}
\centerline{$^{4}$University of Nebraska, Lincoln, Nebraska 68588}
\centerline{$^{5}$New York University, New York, New York 10003}
\centerline{$^{6}$Northwestern University, Evanston, Illinois 60208}
\centerline{$^{7}$State University of New York, Stony Brook,New York 11794.}
}

\date{13 January 1998}
\maketitle

\begin{abstract}
Uncertainties in the NLO calculation of the inclusive jet cross
section due to the choice of renormalization scale, parton
distribution functions and clustering algorithm are explored. These
are found to be similar in size to the current experimental
uncertainties of the measured inclusive jet cross section at D\O\ and
CDF.
\end{abstract}  

\section{Introduction}

 Measurement of the inclusive jet cross section in proton anti-proton
 (\pbarp ) collisions constitutes a strong test of the pdictions of
 perturbative quantum chromodynamics (QCD). Deviations of the
 theoretical cross section from the experimentally observed cross
 section may be evidence for physics beyond the Standard Model. In
 particular, the presence of quark compositeness would enhance the
 cross section at high values of transverse energy (\Et ).

 Recently, the inclusive jet cross section has been measured by the
 CDF~\cite{cdf_ia_inclusive} and D\O ~\cite{d0_inclusive}
 Collaborations with systematic uncertainties ranging from 10$\%$ to
 40$\%$ as function of \Et . With the improvement in the experimental
 accuracy of this measurement, it is worth investigating the accuracy
 of the next to leading order (NLO) QCD
 predictions~\cite{aversa,jetrad,eks}.

 In this paper the uncertainties in the NLO QCD inclusive jet
 calculations are explored using two available programs: {\sc
 Jetrad}~\cite{jetrad} a complete $\mathcal{O}(\als ^{\rm 3})$ event
 generator, and EKS~\cite{eks} a complete $\mathcal{O}$$(\als ^{3})$
 analytical calculation of the inclusive jet cross section. Both
 programs require the selection of a renormalization and factorization
 scale (typically chosen to have the same value, $\mu$), a set of
 parton distribution functions (PDF), and a jet clustering algorithm.

\section{Discussion of the Theoretical Predictions}

 If QCD could be calculated to all orders the results would be
 independent of the choice of renormalization scale. Because the
 inclusive jet cross section has been calculated only to NLO
 ($\mathcal{O}$$(\als ^{3})$), the choice of renormalization scale
 does affect the result. The authors of {\sc Jetrad} have provided
 several choices for the renormalization scale, we have chosen to
 investigate a scale proportional to the \Et\ of the leading jet after
 parton clustering ($\mu = A\Etmax$, where $A$ is a constant typically
 chosen to lie in the range $0.25 \le A \le 2$). The authors of EKS
 prefer an alternative definition of the renormalization scale: the
 \Et\ of each jet in the event ($\mu = A\Etjet$)\footnote{A
 version of EKS that uses the renormalization scale $\mu = A\Etmax$ is
 also available.}. An alternative scheme (available in {\sc Jetrad})
 uses the center-of-mass energy of the two outgoing partons as the
 renormalization scale ($\mu = C\sqrt{\hat{s}} = C\sqrt{x_{1}x_{2}s}$
 where $C$ is a constant chosen to lie in the range $0.25 \le C \le
 1$, $x_{\rm 1} = \sum \Et_{\rm i} e^{\eta_{i}} / \sqrt{s}$, $x_{\rm
 2} = \sum \Et_{\rm i} e^{-\eta_{i}} / \sqrt{s}$, $i = 1 \dots {\rm
 n}$ where n is the number of jets in the event, $\eta = -{\rm
 ln}[{\rm tan}(\theta/2)]$ and $\theta$ is the polar angle relative to
 the proton beam )\footnote{The choice of renormalization scale $\mu =
 C\sqrt{\hat{s}}$ is somewhat unnatural for the inclusive jet cross
 section which is dominated by $t$-channel exchange. It has been
 included to study the effect of an extreme choice of scale and for
 comparison with previous two-jet mass
 analyses~\cite{cdf_dijet_angular}.}. The effect of these scale
 choices on the inclusive jet cross section is discussed.

 The standard Snowmass clustering algorithm ~\cite{snowmass} combines
 two partons into a single jet if they are both within ${\mathcal{R}}
 \equiv \displaystyle{\sqrt{\eta^{\rm 2} + \phi^{\rm 2}}} = 0.7$ of
 their \Et\ weighted center (where $\phi$ is the azimuthal angle). An
 additional constraint on the parton clustering is applied which
 requires that the two partons be closer than ${\mathcal{R}}_{\rm sep}
 = 1.3 \times \mathcal{R}$~\cite{rsep}\footnote{The Snowmass algorithm
 corresponds to ${\mathcal{R}}_{\rm sep} = 2.0 \times
 \mathcal{R}$}. The value of ${\mathcal{R}}_{\rm sep}$ has been chosen
 to match the experimental jet splitting/merging parameters used in
 the jet clustering algorithms~\cite{rsep}. The effect of using
 ${\mathcal{R}}_{\rm sep}$ will be discussed.

 The uncertainty in the calculation of the inclusive jet cross section
 resulting from parton distribution functions will be divided into
 three parts.

 The first is due to the choice of the PDF family. This choice is
 associated with the selection of data used to determine the PDF and
 the functional form used in the fits. The variation of the inclusive
 jet cross section has been studied using a selection of modern PDFs:
 {\sc cteq3m}~\cite{cteq3m}, {\sc cteq4m}, {\sc cteq4hj}~\cite{cteq4m}
 and {\sc mrsa$^\prime$}~\cite{mrsap}.

 The second category of PDF uncertainties results from the value of
 the strong coupling constant (\als ) used in the PDF. Usually a free
 parameter in the PDF fit, \als\ can be fixed to a pre-determined
 value. The effect of \als\ variation on the PDF is examined by using
 the {\sc cteq4a} series in which \als\ is fixed to values ranging
 from 0.110 to 0.122 at $M_Z$. In comparison, $\als (M_{\rm Z})$ =
 0.116 for {\sc cteq4m}.

 Finally, {\sc Jetrad} and EKS use different strategies to evolve the
 PDF in $x$ and $Q^2$ (where $x$ is the momentum fraction carried by
 the parton and $Q$ is the characteristic energy scale of the process,
 typically chosen to be the momentum transfer). The {\sc Jetrad}
 program uses the strategy as implemented by the MRS~\cite{mrsap}
 group for evolving all PDFs ({\sc Jetrad} also uses CTEQ PDFs
 generated using the CTEQ evolution package). EKS implements the PDFs
 by interpolating from a table of values that were generated directly
 from the original PDFs. While these implementations are theoretically
 identical, small differences can be produced by the numerical
 accuracy of the program.

\begin{figure}
\vbox{\centerline{\psfig{figure=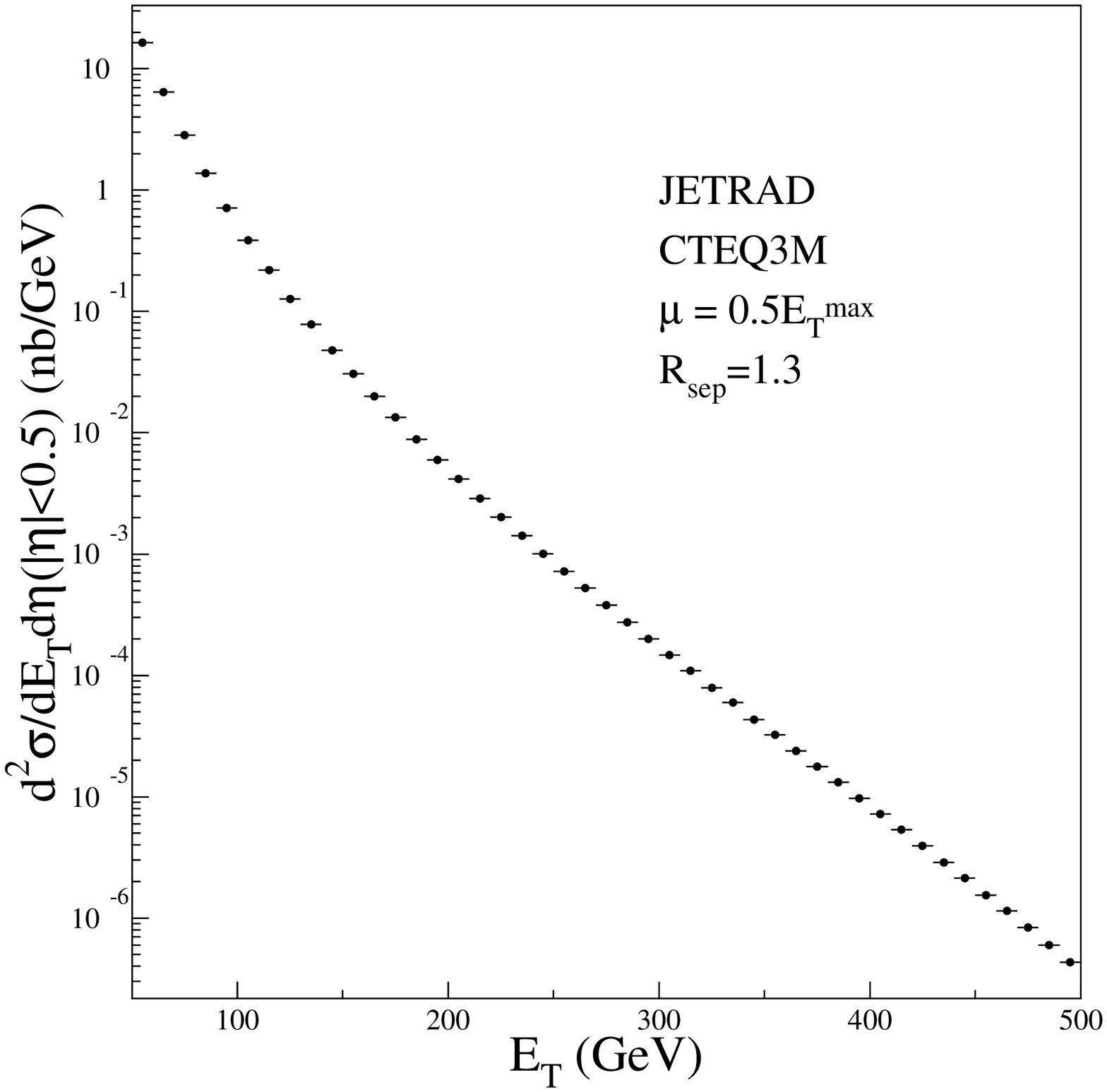,width=6in}}}
\caption{The reference NLO calculation of the inclusive jet cross 
section at $\sqrt{s}$ = 1.8 TeV.  {\sc Jetrad} calculation for the
pseudorapidity range $\modeta < $0.5, $\mu = 0.5\Etmax$,
${\mathcal{R}}_{\rm sep} =$ 1.3 and the {\sc cteq3m} PDF evolved using
the CTEQ method.}
\label{fig_1}
\end{figure}

 The uncertainties will be determined at $\sqrt{s} =$ 1.8 TeV and 630
 GeV and compared to a reference model. For this study, the reference
 model will be the {\sc Jetrad} calculation for the pseudorapidity
 range $\modeta < $0.5, $\mu = 0.5\Etmax$, ${\mathcal{R}}_{\rm sep} =$
 1.3 and the {\sc cteq3m} PDF evolved using the CTEQ method (see
 Fig.~\ref{fig_1}). The comparisons with other theories will be given
 by: $$ R = \left({{{\mbox{\rm Theory}}}\over{\mbox{\rm Reference
 Theory}}} \right){\mbox{\rm{.}}}$$ This ratio is fitted with a third
 degree polynomial yielding a smooth curve. In most cases the ratio
 can be fitted with a resulting $\chi^{2}$ per degree of freedom less
 than one.

\section{Inclusive Jet Cross Section at $\sqrt{s}$ = 1.8 TeV}

\begin{figure}
\vbox{\centerline{\psfig{figure=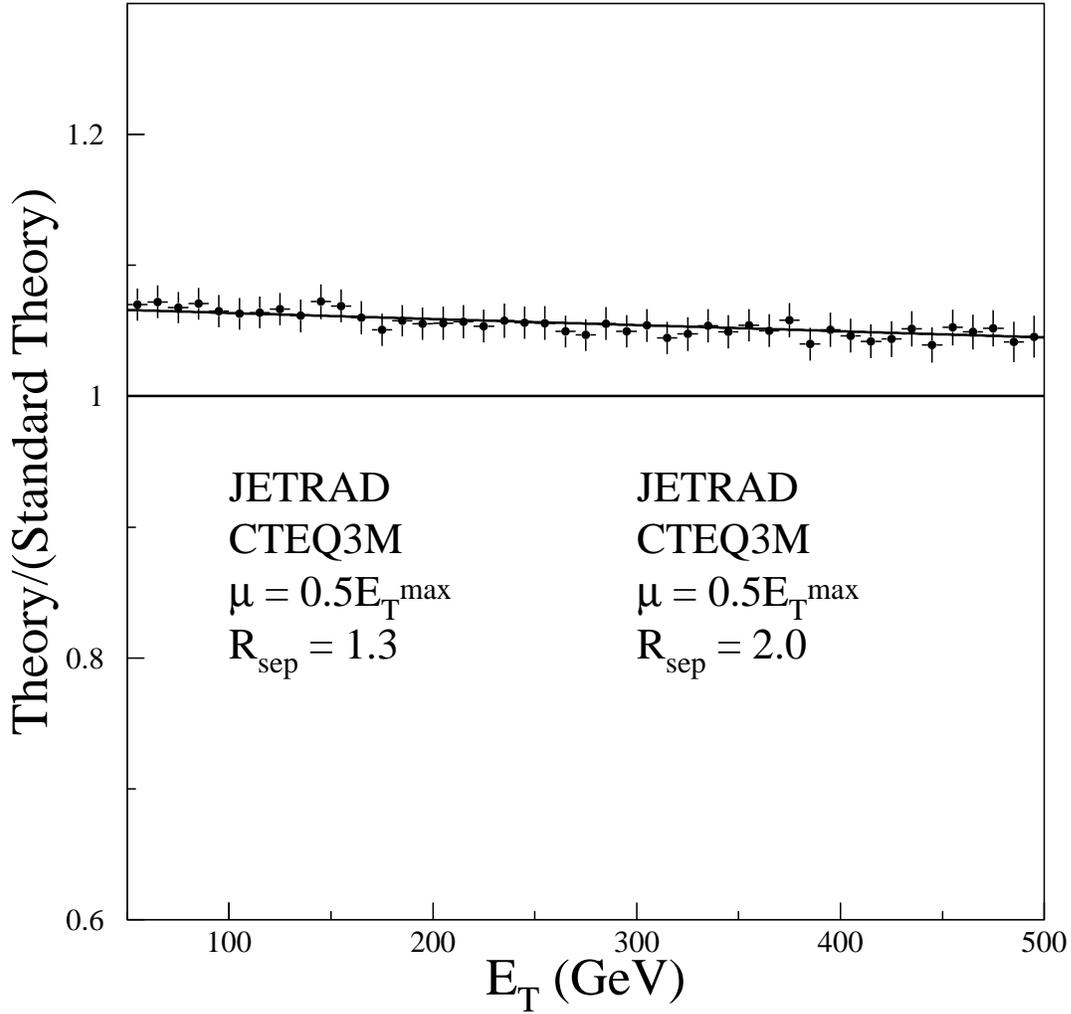,width=6in}}}
\caption{A comparison between the {\sc Jetrad} calculations of 
the inclusive jet cross section with ${\mathcal{R}}_{\rm sep} = 2.0$
and $1.3$ at $\sqrt{s}$ = 1.8 TeV.}
\label{fig_2a}
\end{figure}

 The difference between using the standard Snowmass clustering
 algorithm and the modified algorithm with ${\mathcal{R}}_{\rm sep} =
 1.3$ is shown in Fig.~\ref{fig_2a}. The effect ranges from $8\%$ at
 50 GeV decreasing to $5\%$ at 500 GeV. Because the value of
 ${\mathcal{R}}_{\rm sep}$ is selected to reflect the experimental
 clustering algorithm, the uncertainty resulting from its use is much
 smaller than 5$\%$. A more appropriate variation of the value of
 ${\mathcal{R}}_{\rm sep}$ is from 1.2--1.4 resulting in variations in
 the cross section of less than $1\%$~\cite{rsep}\footnote{The
 uncertainty resulting from the choice of ${\mathcal{R}}_{\rm sep}$
 will not be considered in the remainder of this note.}.

\begin{figure}
\vbox{\centerline
{\psfig{figure=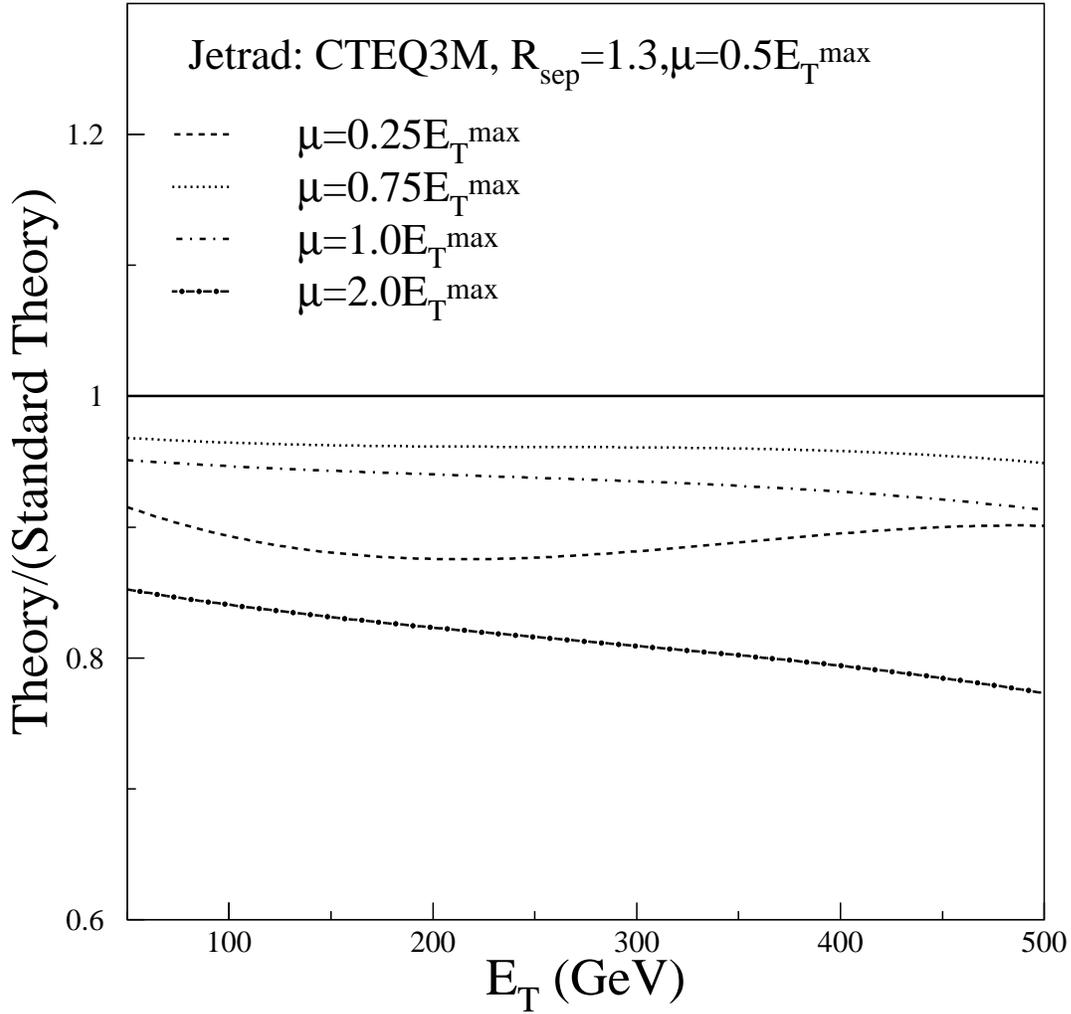,width=6in}}}
\caption{A comparison between the {\sc Jetrad} calculations of 
the inclusive jet cross section with $\mu =$ 0.25, 0.75, 1.0 and $2.0
\Etmax$ compared with $\mu = 0.5\Etmax$ at $\sqrt{s}$ = 1.8 TeV.}
\label{fig_3}
\end{figure}

 The effect on the cross section due to the choice of the
 renormalization scale was studied by using {\sc Jetrad} with several
 different values of $\mu$. First, the cross section was calculated
 using $\mu =$ 0.25, 0.75, 1.0 and 2.0 $\Etmax$. These are compared to
 the cross section with $\mu =$ 0.5 $\Etmax$ in Fig.~\ref{fig_3}. The
 cross section is largest\footnote{The inclusive jet cross section at
 $\mu =$ 0.5 $\Etmax$ is a maximum since this is a point of minimum
 sensitivity for the calculation, see~\cite{jets_and_partons} for a
 discussion.} for $\mu = 0.5\Etmax$ and is reduced by 5--10$\%$ with
 some \Et\ dependence for $\mu = 1.0 \Etmax$ and $0.75 \Etmax$. The
 cross section is approximately 10$\%$ below the reference model for
 $\mu = 0.25\Etmax$ and 15--20$\%$ below the reference model for $\mu
 = 2.0\Etmax$ with some \Et\ dependence.

\begin{figure}
\vbox{\centerline
{\psfig{figure=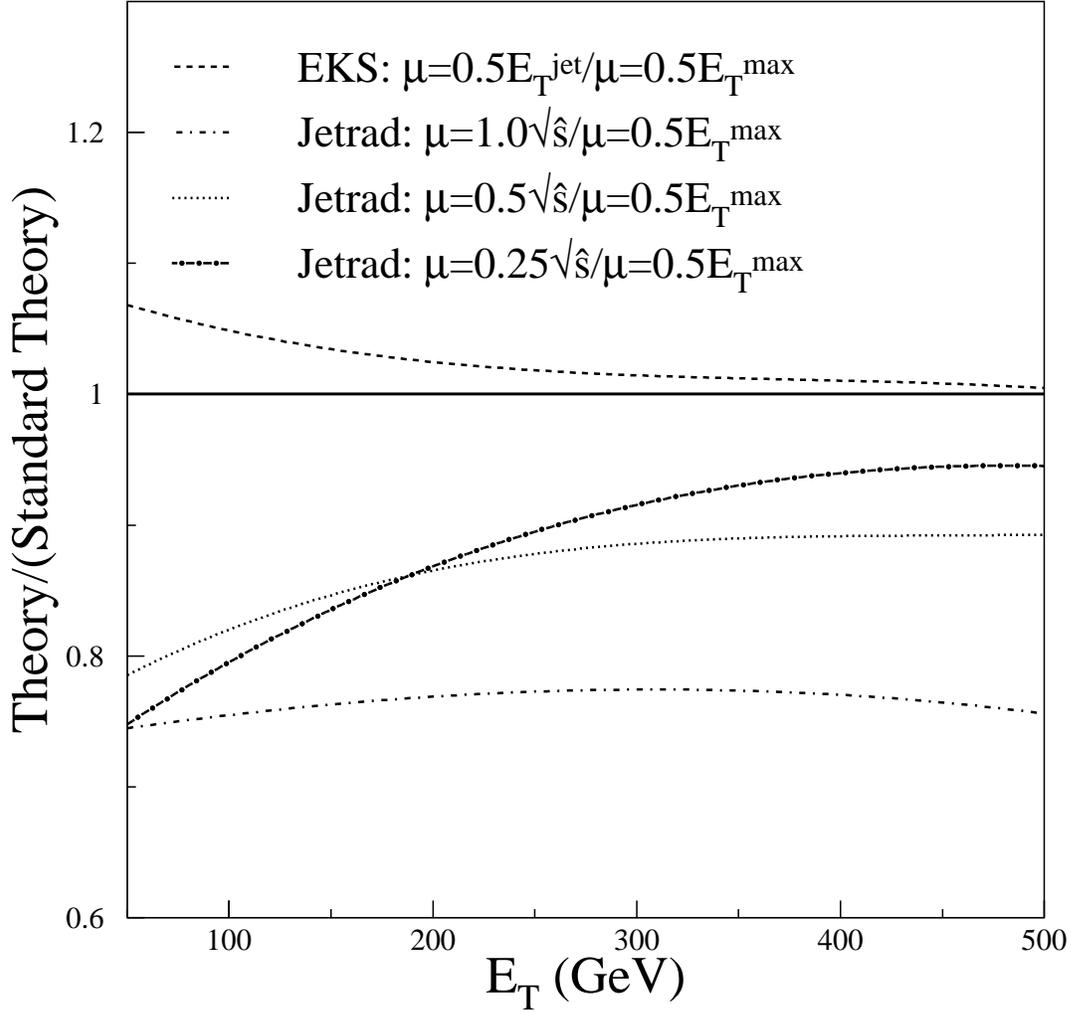,width=6in}}}
\caption{A comparison between the EKS calculation of the inclusive jet 
cross section with $\mu = 0.5\Etjet$ and $\mu =
0.5\Etmax$. Also shown is a comparison between the {\sc
Jetrad} calculations of the inclusive jet cross section with $\mu =$
0.25, 0.5 and $1.0\sqrt{\hat{s}}$ compared with $\mu = 0.5
\Etmax$ at $\sqrt{s}$ = 1.8 TeV.}
\label{fig_4}
\end{figure}

 Figure~\ref{fig_4} compares the predictions for alternative choices
 of $\mu$, $\Etjet$ and $\sqrt{\hat{s}}$. The choice $\mu =
 0.5\Etjet$ (calculated with EKS\footnote{{\sc Jetrad} does not
 implement $\mu = A\Etjet$.})  is compared to $\mu =
 0.5\Etmax$ and shows a $5\%$ difference at an \Et\ of 50 GeV
 dropping to less than 1$\%$ at 500 GeV. Also shown in
 Fig.~\ref{fig_4} is a comparison of the {\sc Jetrad} calculations for
 $\mu =$ 0.25, 0.5, 1.0 $\sqrt{\hat{s}}$ compared with $\mu =$ 0.5
 $\Etmax$. The effect is approximately 20--25$\%$ at 50 GeV
 decreasing to 10$\%$ at 500 GeV with a strong dependence on the
 choice of scale used. In summary, there are only small differences
 between the scale choices of $\Etjet$ and $\Etmax$ but
 large differences when the scale is changed from \Et\ to
 $\sqrt{\hat{s}}$.

\begin{figure}
\vbox{\centerline{\psfig{figure=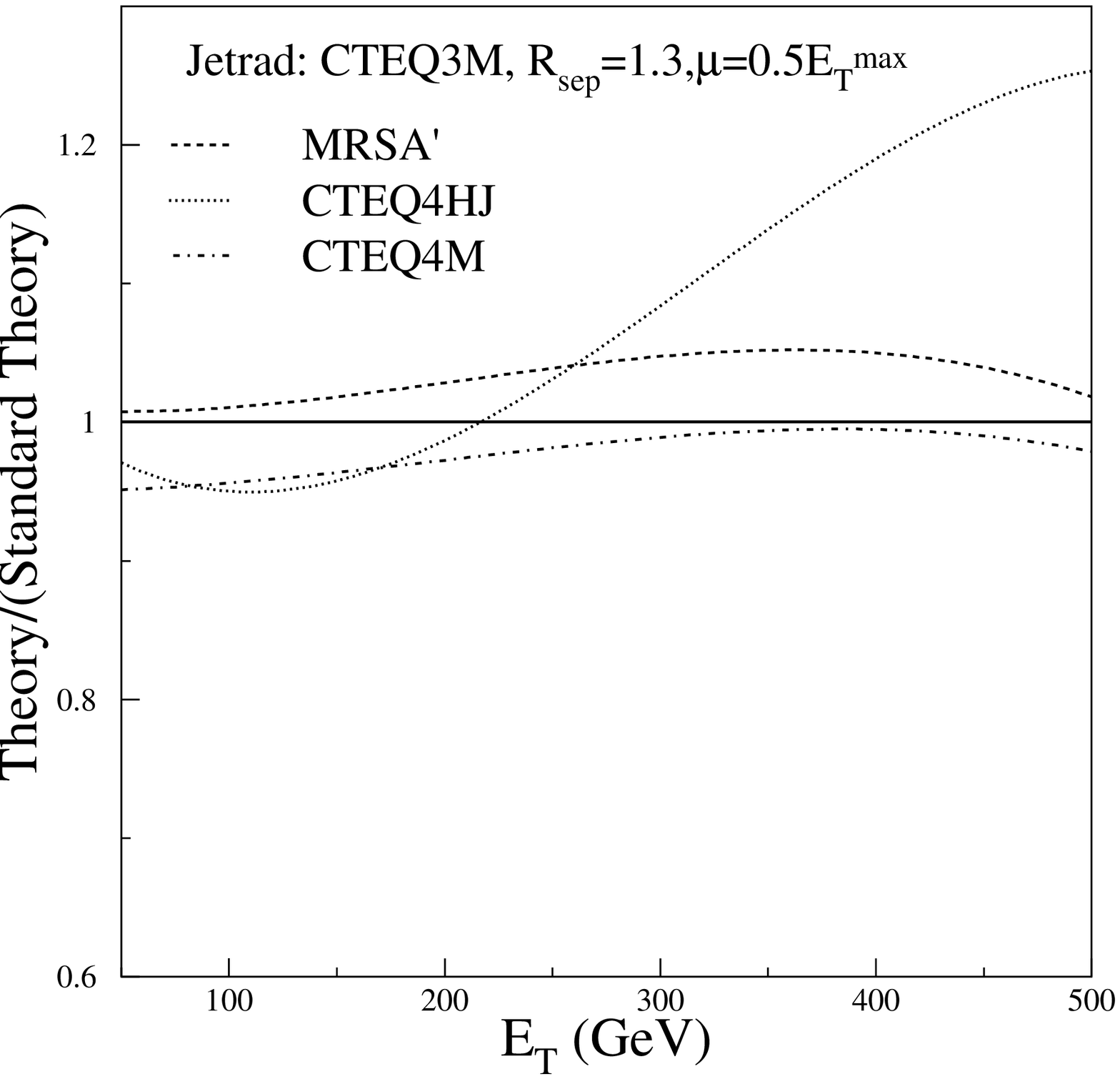,width=6in}}}
\caption{A comparison between the {\sc
Jetrad} calculations of the inclusive jet cross section with the {\sc
cteq4m, cteq4hj} and the {\sc mrsa$^\prime$} PDF compared with the
calculation using {\sc cteq3m} at $\sqrt{s}$ = 1.8 TeV.}
\label{fig_5}
\end{figure}

\begin{figure}
\vbox{\centerline{\psfig{figure=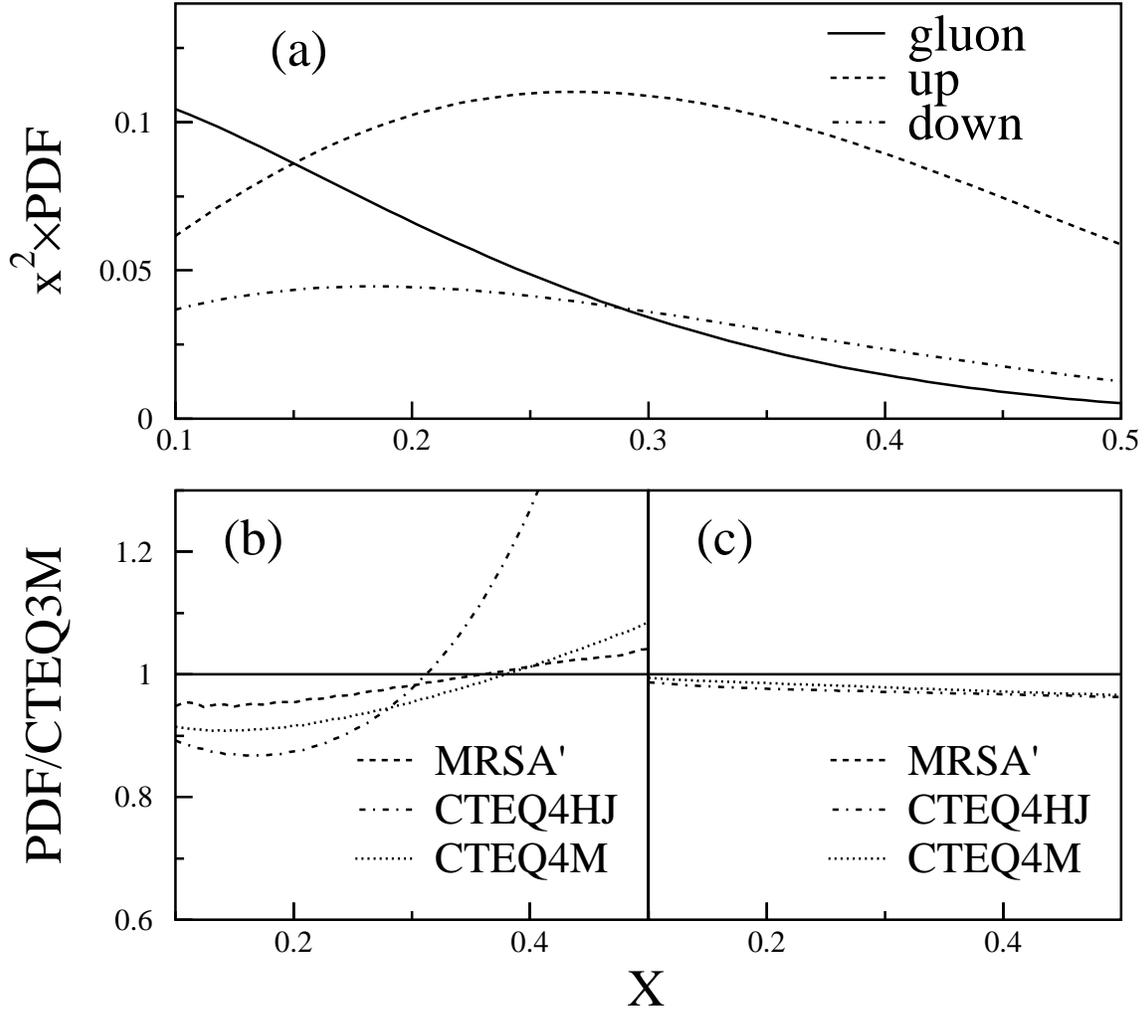,width=6in}}}
\caption{A comparison of the PDFs used in this analysis. (a) The 
{\sc cteq3m} PDF for $Q=100$ GeV (the momentum transfer) as a function
of the momentum fraction carried by the parton ($x$). (b) A comparison
between the gluon distributions from the {\sc cteq4m, cteq4hj} and the
{\sc mrsa$^\prime$} PDFs compared with the {\sc cteq3m} PDF. (c) A
comparison between the up quark distributions from the {\sc cteq4m,
cteq4hj} and the {\sc mrsa$^\prime$} PDFs compared with the {\sc
cteq3m} PDF.}
\label{fig_5a}
\end{figure}

 The effect of different PDF choices is depicted in
 Fig.~\ref{fig_5}. The cross section calculated with {\sc cteq4m} is a
 few percent lower than the cross section calculated using {\sc
 cteq3m}. These differences result from a change in the
 parameterization used to model the gluon distribution\footnote{{\sc
 cteq3m} uses a more restrictive gluon parameterization:
 $G(x,Q_{0})=A_{0} x^{A_{1}}(1-x)^{A_{2}}(1+A_{3} x)$ while {\sc
 cteq4m} uses: $G(x,Q_{0})=A_{0} x^{A_{1}}(1-x)^{A_{2}}(1+A_{3}
 x^{A_{4}})$~\cite{cteq4m} } of the {\sc cteq4m} PDF and the inclusion
 of additional data sets. These data sets include more precise deep
 inelastic scattering (DIS) data from NMC~\cite{NMC} and
 HERA~\cite{H1,ZEUS}, the inclusive jet cross section measured by
 CDF~\cite{cdf_ia_inclusive,cdf_1b_inclusive} and the preliminary
 inclusive jet cross section measured by D\O\
 \cite{old_d0_inclusive}. These changes lead to a change in the
 optimal value of \als\ from 0.112 to 0.116~\cite{cteq4m}. The {\sc
 cteq4hj} PDF, which emphasizes recent Tevatron jet data to constrain
 the gluon distribution, shows a decrease in the cross section at low
 \Et\ of approximately 5$\%$ and an increase in the cross section at
 500 GeV of approximately 25$\%$. The final comparison is made using
 {\sc mrsa$^\prime$}, which uses a slightly different parameterization
 and input data to {\sc cteq3m}. This results in cross sections that
 are similar at high and low \Et\ and approximately 5$\%$ higher at
 300 GeV. These differences are caused by the gluon distributions of
 the PDFs (see Fig.~\ref{fig_5a}). Note that the variations in the
 cross section calculations due to the choice of PDF are limited by
 the similarity of the parameterizations used to model the gluon
 distributions in the PDFs (which are not well constrained by
 experiment). The choice of an alternative parameterization of the
 gluon PDF could lead to larger uncertainties.

\begin{figure}
\vbox{\centerline
{\psfig{figure=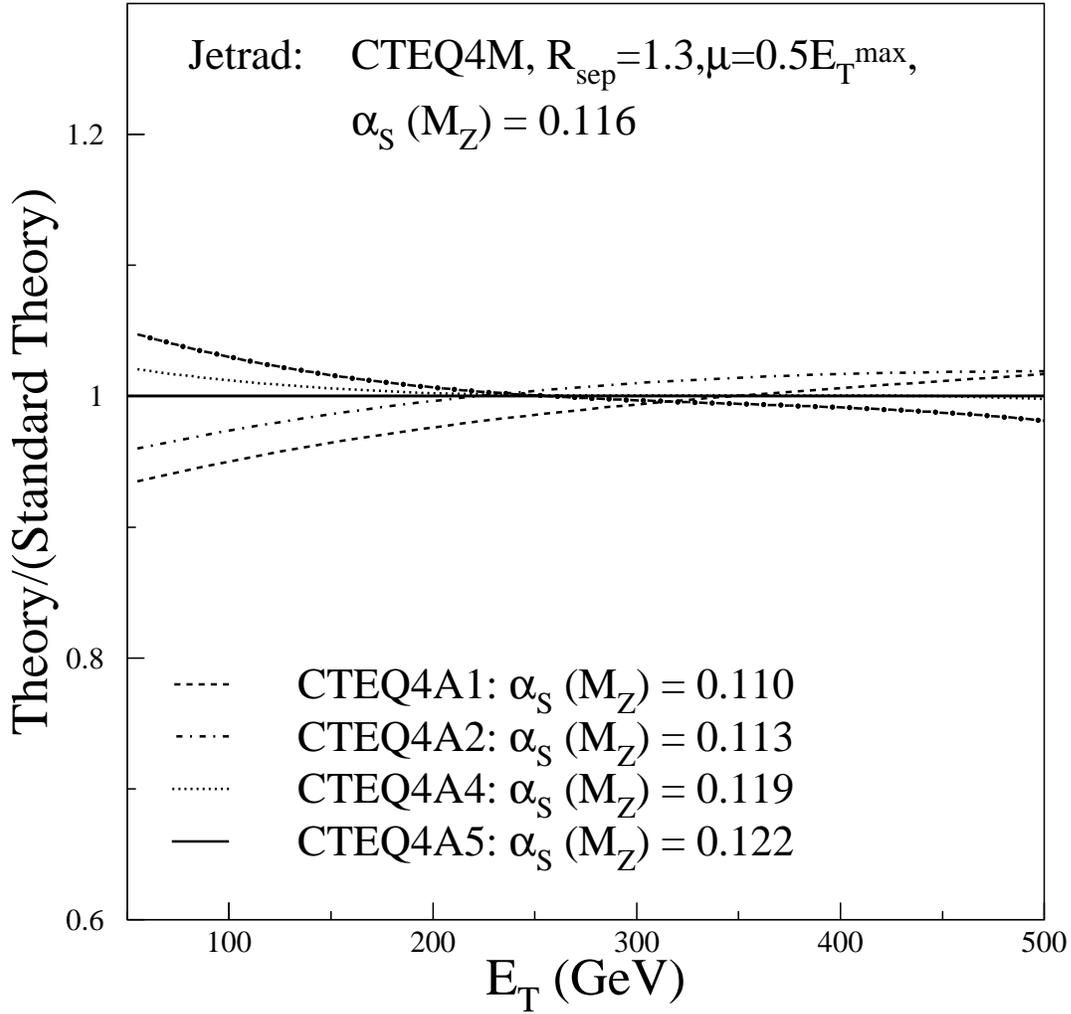,width=6in}}}
\caption{A comparison between the {\sc Jetrad} calculations of the 
inclusive jet cross section with the {\sc cteq4a} series of PDFs
compared with the calculation using {\sc cteq4m} at $\sqrt{s}$ = 1.8
TeV.}
\label{fig_6}
\end{figure}

 The cross section change due to the variation of \als\ in the PDF is
 shown in Fig.~\ref{fig_6}.  Approximately $\pm5\%$ changes are seen
 at low \Et\ which diminish as the \Et\ increases. The uncertainty in
 the PDF due to the choice of \als\ is significantly smaller than the
 uncertainty due to the gluon distributions.

\begin{figure}
\vbox{\centerline{\psfig{figure=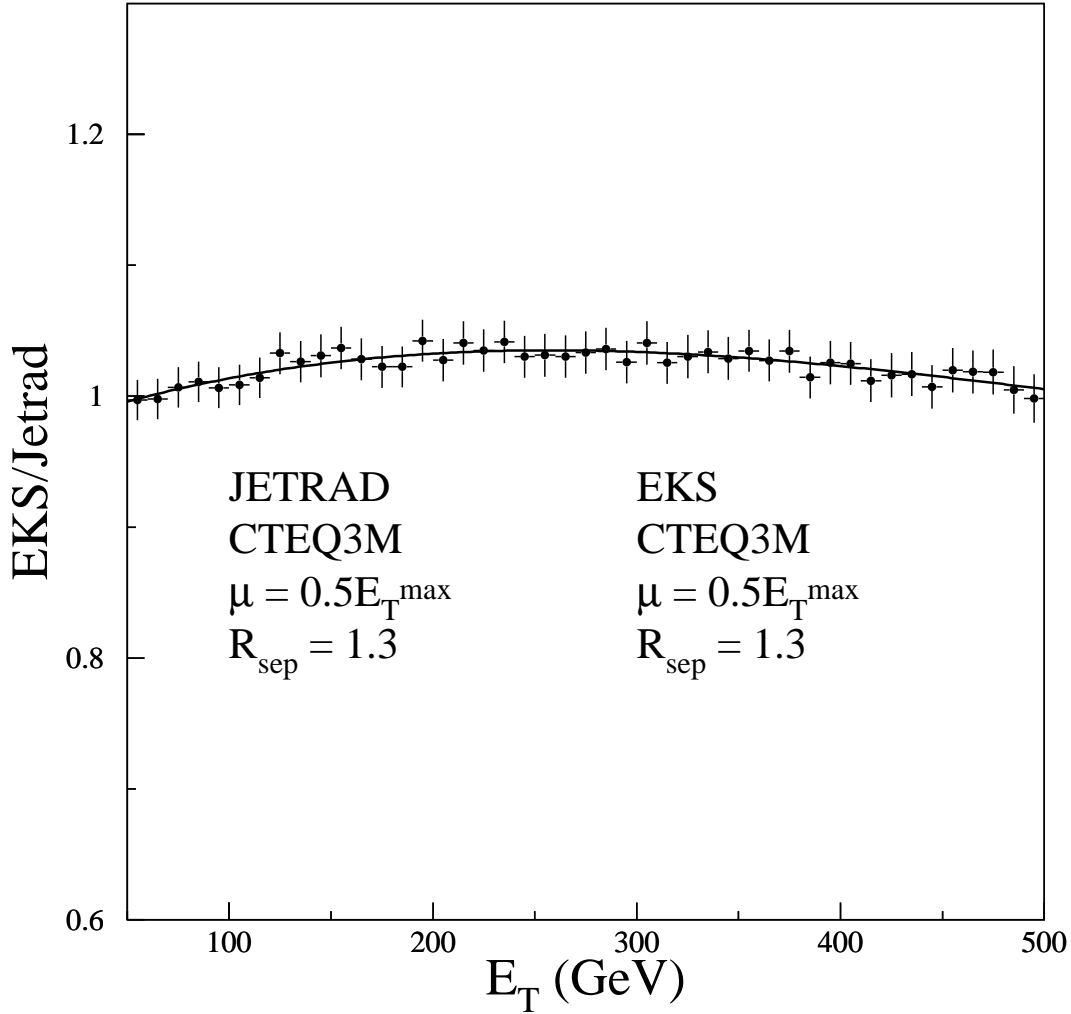,width=6in}}}
\caption{A comparison between the {\sc Jetrad} and EKS calculations of 
the inclusive jet cross section at $\sqrt{s}$ = 1.8 TeV. The theory
parameters are $\modeta < $0.5, $\mu = 0.5\Etmax$,
${\mathcal{R}}_{\rm sep} =$ 1.3 and the {\sc cteq3m} PDF. The fit is
to a  third degree polynomial.}
\label{fig_2}
\end{figure}

\begin{figure}
\vbox{\centerline{\psfig{figure=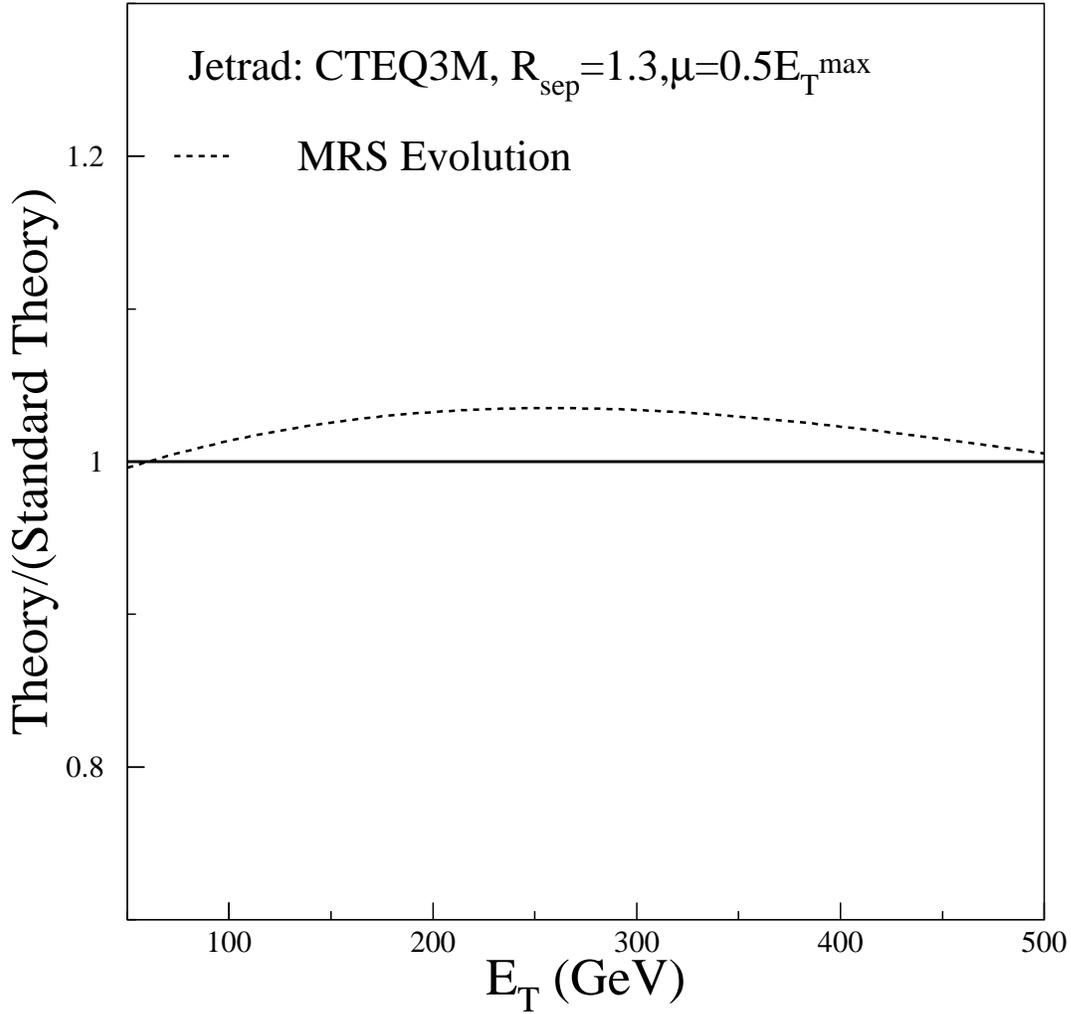,width=6in}}}
\caption{A comparison between {\sc Jetrad} calculation of the inclusive 
jet cross section with the {\sc cteq3m} PDF using the MRS and the CTEQ
evolution packages.}
\label{fig_7}
\end{figure}

 A comparison can be made between the EKS and {\sc Jetrad}
 calculations for $\modeta < $0.5 using $\mu = 0.5\Etmax$,
 ${\mathcal{R}}_{\rm sep} =$ 1.3 and the {\sc cteq3m} PDF.  As shown
 in Fig.~\ref{fig_2}, the two calculations differ at a level of
 2-3$\%$ with some dependence on the \Et . This variation is due to
 the different evolution in $x$ and $Q^2$ used by the two programs.
 Figure~\ref{fig_7} shows the variation in the {\sc Jetrad}
 predictions due to the different evolution methods.  These
 differences lead to $5\%$ difference in the cross section with some
 \Et\ dependence, similar to the differences between the EKS and {\sc
 Jetrad} programs.

\section{Inclusive Jet Cross Section at $\sqrt{s}$ = 630 GeV}

\begin{figure}
\vbox{\centerline{\psfig{figure=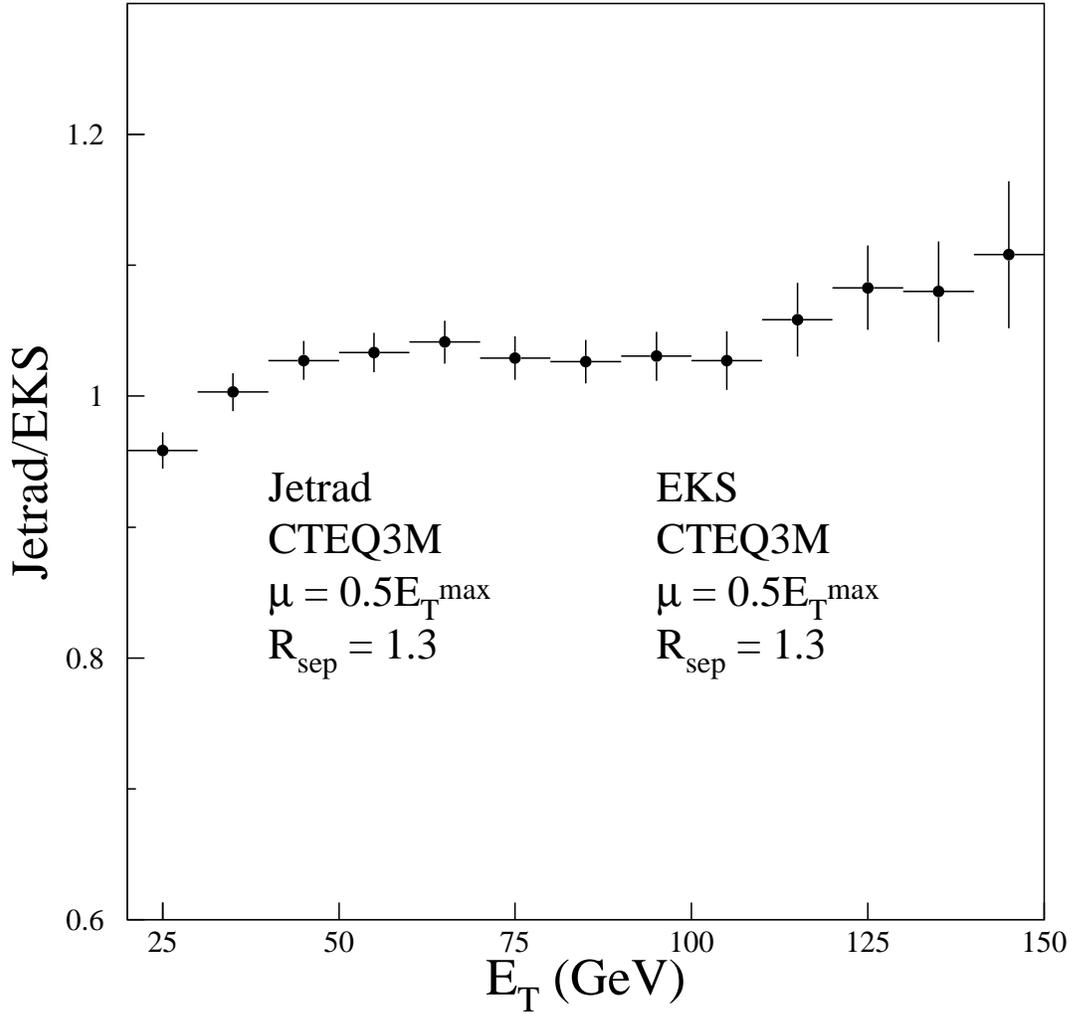,width=6in}}}
\caption{A comparison between the {\sc Jetrad} and EKS calculations of 
the inclusive jet cross section at $\sqrt{s}$ = 630 GeV. The theory
parameters are $\modeta < $0.5, $\mu = 0.5\Etmax$,
${\mathcal{R}}_{\rm sep} =$ 1.3 and the {\sc cteq3m} PDF.
}
\label{fig_8}
\end{figure}

 The studies described in the previous section were repeated for
 $\sqrt{s} = $ 630 GeV. The comparison between {\sc Jetrad} and EKS
 with $\mu = A\Etmax$ is given in Fig.~\ref{fig_8}. EKS
 produces a cross section that is 5$\%$ lower than the {\sc Jetrad}
 cross section at 20 GeV and 10$\%$ higher at 150 GeV. These
 uncertainties cannot be fully explained by the difference in choice
 of PDF evolution.

\begin{figure}
\vbox{\centerline{\psfig{figure=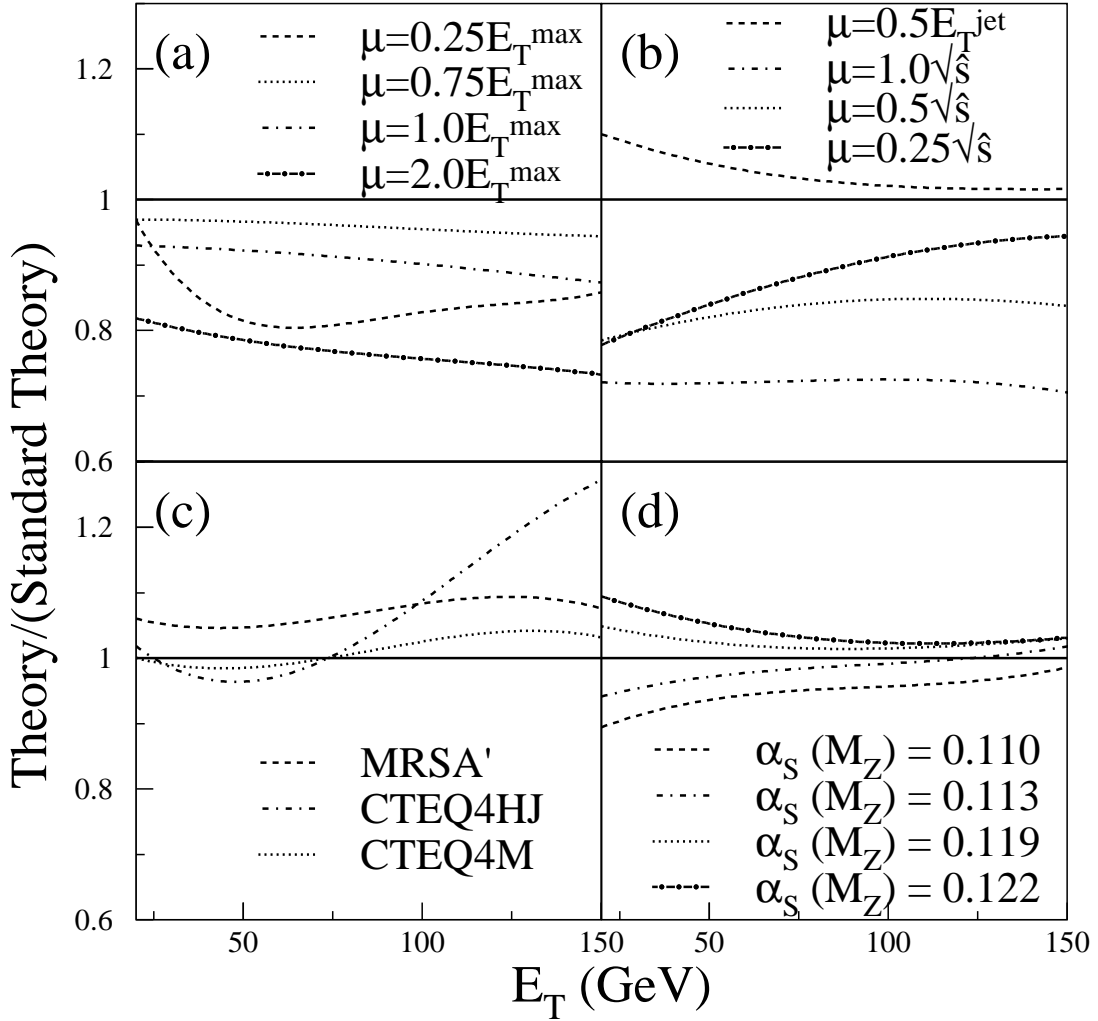,width=6in}}}
\caption{The deviations of the inclusive jet cross section at 
$\sqrt{s}$ = 630 GeV.  (a) A comparison between the {\sc Jetrad}
calculations with $\mu =$ 0.25, 0.75, 1.0 and $2.0
\Etmax$ compared with $\mu = 0.5 \Etmax$. (b) A 
comparison between the EKS calculation with $\mu = 0.5\Etjet$
and $\mu = 0.5\Etmax$. Also shown is a comparison between the
{\sc Jetrad} calculations with $\mu =$ 0.25, 0.5 and $1.0
\sqrt{\hat{s}}$ compared with $\mu = 0.5 \Etmax$. 
(c) A comparison between the {\sc Jetrad} calculations with the
{\sc cteq4m, cteq4hj} and the {\sc mrsa$^\prime$} PDF compared with
the calculation using {\sc cteq3m}. (d) A comparison between the {\sc
Jetrad} calculations with the {\sc cteq4a} series of PDFs compared
with the calculation using {\sc cteq4m}.  }
\label{fig_9}
\end{figure}

 The variations in the NLO calculations of the cross section due to
 the choice of renormalization scale and PDF are given in
 Fig.~\ref{fig_9}. These variations are slightly larger than those
 observed at $\sqrt{s}$ = 1.8 TeV.

\section{The ratio of the Inclusive Jet Cross Sections at $\sqrt{s}$ = 
         1.8 TeV and 630 GeV}

 Theoretical uncertainties in the NLO QCD predictions should be
 reduced in the ratios of the inclusive jet cross sections at
 $\sqrt{s}$ = 630 GeV and 1.8 TeV as a function of the dimensionless
 quantity $X_{\rm T} = 2\Et/{\sqrt{s}}$.

\begin{figure}
\vbox{\centerline{\psfig{figure=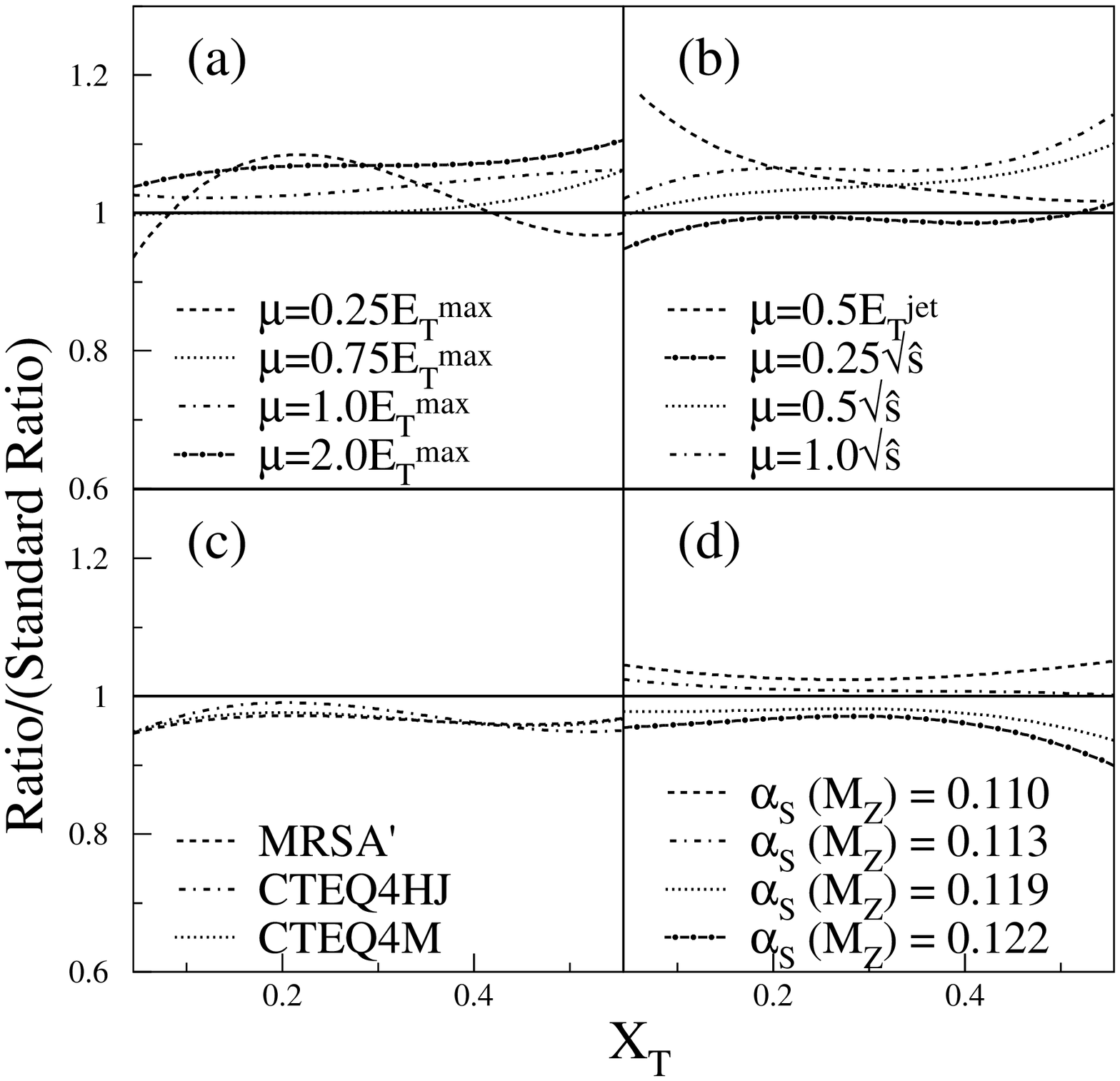,width=6in}}}
\caption{The deviations of the ratio of inclusive jet cross sections 
at $\sqrt{s}$ = 1800 GeV and 630 GeV.  (a) A comparison between the
{\sc Jetrad} calculations with $\mu =$ 0.25, 0.75, 1.0 and $ 2.0
\Etmax$ compared with $\mu = 0.5 
\Etmax$. (b) A comparison between the EKS calculation 
with $\mu = 0.5\Etjet$ and $\mu = 0.5\Etmax$. Also shown is a
comparison between the {\sc Jetrad} calculations with $\mu =$ 0.25,
0.5 and $ 1.0 \sqrt{\hat{s}}$ compared with $\mu = 0.5\Etmax$. (c) A
comparison between the {\sc Jetrad} calculations with the {\sc cteq4m,
cteq4hj} and the {\sc mrsa$^\prime$} PDF compared with the calculation
using {\sc cteq3m}. (d) A comparison between the {\sc Jetrad}
calculations with the {\sc cteq4a} series of PDFs compared with the
calculation using {\sc cteq4m}.  }
\label{fig_10}
\end{figure}

 The variations in the ratio of the cross sections due to the choice
 of renormalization scale is approximately 15$\%$ with some dependence
 on $X_{\rm T}$ (see Fig.~\ref{fig_10} (a) and (b)). The uncertainty
 due to the choice of PDF is only a few percent (Fig.~\ref{fig_10}
 (c)) which compares to an uncertainty of up to 25$\%$ in the
 individual cross sections. The difference due to the variation of
 \als\ is not reduced by measuring the ratio of the cross sections and
 is still at the 5$\%$ level (Fig.~\ref{fig_10} (d)).
 
\section{Conclusion}

 The inclusive jet cross section, predicted using the available NLO
 programs, has significant uncertainties due to the choice of
 renormalization scale and PDF. The overall variation in the cross
 section can be as large as $\pm30\%$. Except for alternative
 implementation of PDFs and the evolution strategies used, {\sc
 Jetrad} and EKS appear to be identical.

 The ratio of the inclusive jet cross sections at $\sqrt{s}$ = 630 GeV
 and 1.8 TeV as a function of $X_T$ has uncertainties of approximately
 10--20$\%$, which is much smaller than the variation of the cross
 sections.

 Before the inclusive jet cross section can be used to test QCD or
 search for New Phenomena\footnote{This result does not not eliminate
 the possibility of observing new physics that produce effects on the
 inclusive jet cross section that are significantly larger than the
 theoretical uncertainties.}, the theoretical predictions must
 improve. Most feasibly through improved measurement of the gluon
 distributions.

 Motivation for this work work arose as a result of discussions within
 the D\O\ QCD group and in discussions between that group, W.~Giele,
 D.~Soper, and the CDF QCD group. We thank the staffs at Fermilab and
 collaborating institutions for their contributions to this work, and
 acknowledge support from the Department of Energy and National
 Science Foundation (U.S.A.).

\end{document}